\documentclass[%
 reprint,
superscriptaddress,
 amsmath,amssymb,
 aps,
pre,
]{revtex4}

\usepackage{graphicx}
\usepackage{dcolumn}
\usepackage{bm}
\usepackage{hyperref}
\usepackage{amssymb}   
\usepackage{amsmath}
\usepackage{epstopdf}

\usepackage{xcolor}
\newcommand{\blue}[1]{\textcolor{blue}{#1}}

\begin{document}

\title{On the design of particle filters inspired by animal noses} 

\author{Jisoo Yuk}
\affiliation{Department of Biological and Environmental Engineering, Cornell University, Ithaca, NY 14850, United States}
\author{Aneek Chakraborty} 
\affiliation{Department of Mechanical Engineering, Jadavpur University, Kolkata, WB 700032, India}
\author{Shyuan Cheng}
\affiliation{Department of Mechanical Science and Engineering, Urbana, IL 61820, United States}
\author{Chun-I Chung}
\affiliation{Department of Mechanical Science and Engineering, Urbana, IL 61820, United States}

\author{Ashley Jorgensen} 
\affiliation{Department of Mechanical Engineering, South Dakota State University, Brookings, SD 57007, United States}

\author{Saikat Basu}  \email{Saikat.Basu@sdstate.edu}
\affiliation{Department of Mechanical Engineering, South Dakota State University, Brookings, SD 57007, United States}

\author{Leonardo P. Chamorro} \email{lpchamo@illinois.edu}
\affiliation{Department of Mechanical Science and Engineering, Urbana, IL 61820, United States}

\author{Sunghwan Jung} \email{sunnyjsh@cornell.edu}
\affiliation{Department of Biological and Environmental Engineering, Cornell University, Ithaca, NY 14850, United States}

\date{\today}

\begin{abstract}
Passive filtering is a common strategy used to reduce airborne disease transmission and particulate contaminants in buildings and individual covers.  The engineering of high-performance filters with relatively low flow resistance but high virus- or particle-blocking efficiency is a nontrivial problem of paramount relevance, as evidenced in the variety of industrial filtration systems and the worldwide use of face masks. In this case, standard N95-level covers have high virus-blocking efficiency, but they can cause breathing discomfort. Next-generation industrial filters and masks should retain sufficiently small droplets and aerosols while having low resistance. We introduce a novel 3D printable particle filter inspired by animals' complex nasal anatomy. Unlike standard random-media-based filters, the proposed concept relies on equally spaced channels with tortuous airflow paths. These two strategies induce distinct effects: a reduced resistance and a high likelihood of particle trapping by altering their trajectories with tortuous paths and induced local flow instability. The structures are tested for pressure drop and particle filtering efficiency over a wide range of airflow rates. We have also cross-validated the observed efficiency through numerical simulations. The designed filters exhibit a lower pressure drop than the commercial mask and air filters (N95, surgical, and high-efficiency particulate air (HEPA)). The concept provides a new approach to developing scalable, flexible, high-efficiency air filters for various engineering applications.
\end{abstract}



\maketitle


\section*{Introduction} \label{sec:intro} 

The control of airborne disease transmission and contaminants has been of central priority across various scientific and engineering disciplines in the last decades. Despite the technological advancements, the emergence of new material designs \citep{souzandeh2019towards} and a better understanding of the particle-capture processes, passive filtering remains a widely used method to address multiple needs across scales and operational requirements. Most standard filters on the market rely on porous membranes  and fibrous structures \citep{liu2015efficient,li2017novel}. However, high filtration efficiency requires sufficiently small pores and holes, which may induce excessive pressure drop and, eventually, clogging \citep{li2017novel}. Fiber-based filters consist of random arrangements of fiber networks of different composition \citep{jung2020advanced}, where interception, inertial impact, diffusion, and gravitational settling \citep{zhu2017electrospun} are the main mechanisms to capture particles.

Physical experiments, numerical simulations, and mathematical modeling \citep{ brown1993air, kang2019modeling} have been extensively used to characterize the filtration efficiency, which provide new insight into an optimal design and underlying working principles.   Analysis of pores in a statistically homogeneous, isotropic, or anisotropic grid of fibers by Castro and Ostoja-Starzewski \cite{Castro} and Bliss \textit{et al.} \cite{Bliss} showed that the probability of particulate retention approaches to unity only in the limit of infinitely dense systems. They pointed out that there are always comparatively large pores where particles can percolate in any randomly-structured filters.

Fiber-based filters are ubiquitous in heating, ventilation, air conditioning (HVAC) systems, industrial processes and personal respirators.  Both solid and liquid aerosols, which are primary targets in designing such filters, can severely impact human health. For instance, particulate matters of sizes smaller than 2.5 $\mu$m can reach the thoracic region and the circulatory system and cause respiratory and cardiovascular diseases \cite{feng2016health}. In the particular case of droplets (typically $\gtrsim 5 ~\mu$m) and aerosols ($\lesssim 5~\mu$m) generated via coughing, sneezing, and speech have also been shown to transport various pathogens \cite{bazant2021guideline, mittal2020flow}. In general, the hazardous size range for inhaled droplets and aerosols that may trigger the initial infection hot-spot along the upper airway is roughly 2 -- 20 $\mu$m \cite{basu2021scirep}. Although increasing a filter's particle-capture efficiency could be modulated by, e.g., smaller inlet holes and multiple layers of fibers, the process is still at the expense of higher power requirements to drive air transport. Capturing efficiency and associated pressure drop are central factors in designing a filter \cite{chaudhuri2019pressure}.  The latter is linearly related in Stokes flows and exhibits a quadratic relationship at sufficiently high Reynolds numbers.

Animals are known to have a sense of smell much better than humans; e.g., approximately $50$ times for mice, $200$ times for dogs, and $700$ times for pigs \cite{laska2017human}. Most of those with highly evolved olfactory systems have tortuous air pathways along the nasal cavity with long and curved turbinates that split and stretch the inhaled air's streamlines.  The multiscale morphological structures and other biological features help capture tiny particles from the inhaled air onto the olfactory epithelium \cite{spencer2021sniffing,zwicker2018physical}. Airpath twists induce secondary flows, including the so-called Dean's pattern. Such motions increase the residence time of particles traveling into the nose system and the likelihood of capture, resulting in highly efficient filtering at a relatively low-pressure drop. 

Inspired by this biological mechanism, we have developed 3D-printable filters that employ distinct features of animals' nasal structures. Our passive filter concept comprises a parallel array of conduits with organized, simplified tortuosity set by analyzing the characteristic dimensions of animals' nasal cavities through CT scan images. The pressure drop and particle filtering efficiency of the proposed concept are measured through experiments and simulations and are compared with their commercial counterparts. It is also worth pointing out that fiber-baser commercial filters may change filtration efficiency and pressure drop when exposed to solid-liquid and liquid-solid aerosols \cite{gac2016consecutive}; however, the performance of the proposed concept is agnostic to the constitution of particulates and droplets.


\section*{Results}

\begin{figure*}[t]
\centering
\includegraphics[width=1\textwidth]{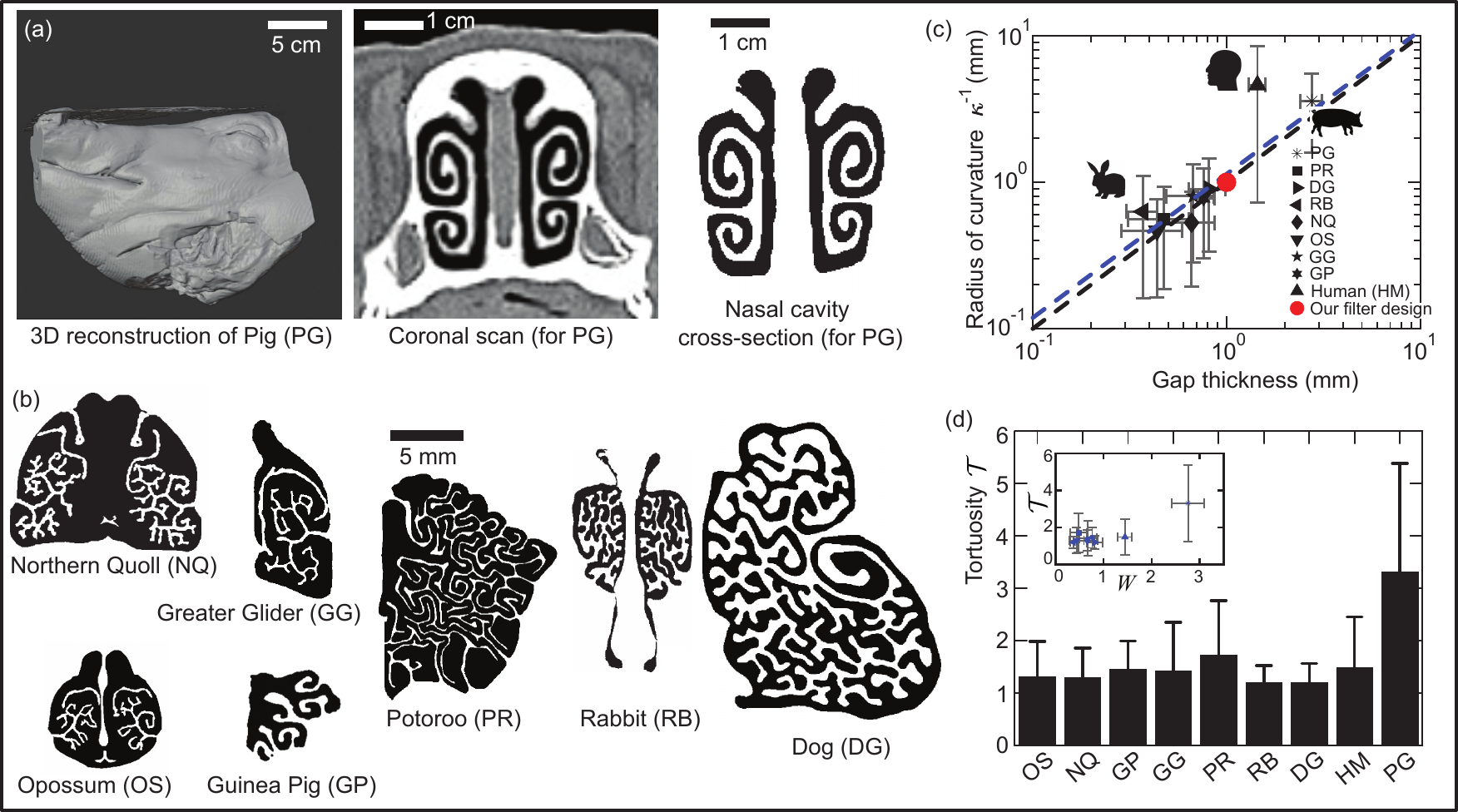} 
\caption{(a) 3D reconstructed model and coronal view from computed tomography (CT) imaging of an adult pig. The image on the far right shows only the nasal airspace (in black), extracted from the CT scan.  (b) Coronal views of the nasal structure of various animals (Northern quoll,  opossum,  greater glider, and potoroo images are from \cite{macrini2012comparative}. Guinea pig image is from \cite{rodgers2012cavia}. Rabbit image is from  \cite{casteleyn2010nalt}. Dog image is from \cite{craven2009development}.)  (c) Relationship between the gap thickness and the radius of curvature measured in 9 animals' skulls; the black-dotted line of unity slope is included for reference. (d) Measured tortuosity from the various animals' nasal cavity structures. Here, the $x$-axis is defined by the animal weight, and the light red region indicates the tortuosity range of the proposed filter designs.}
\label{Fig_animal}
\end{figure*}

\begin{figure*}[t]
\centering
\includegraphics[width=1\textwidth]{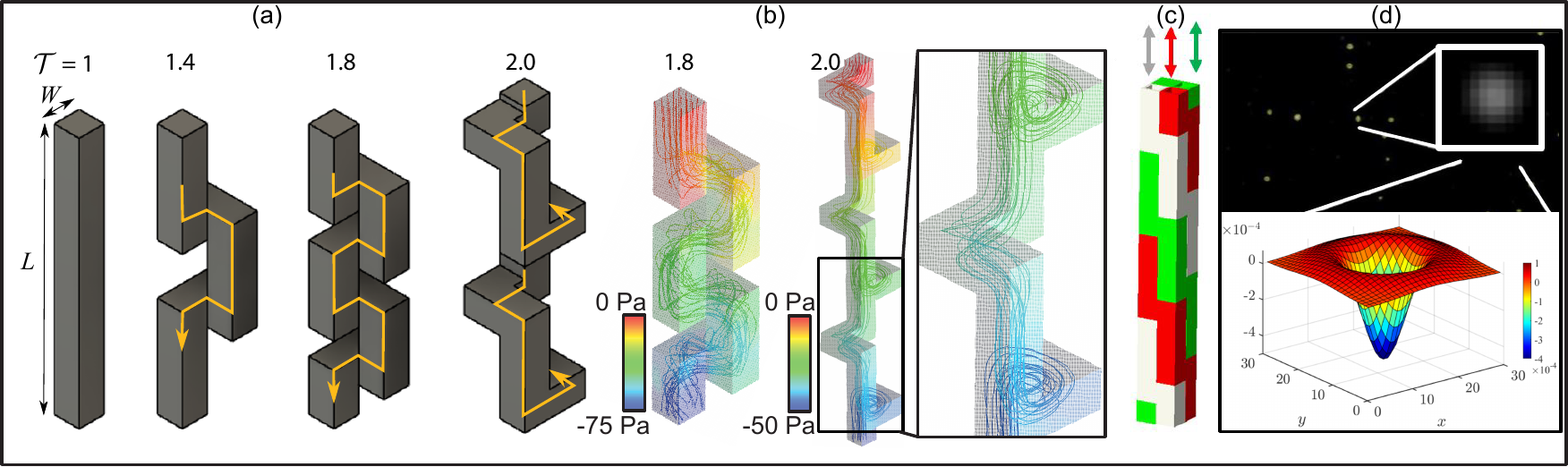}
\caption{(a) Schematics of filter conduits with different tortuosity, ${\cal T}$; b) examples of simulated streamlines in a single filter pathway highlighting complex mean flow trajectories; c) example of 'packed' conduits for reduced pressure drop; d) illustration of experimental particle characterization using Gaussian kernel. }
\label{tau}
\end{figure*}

\begin{figure}[h]
\centering
\includegraphics[width=0.5\textwidth]{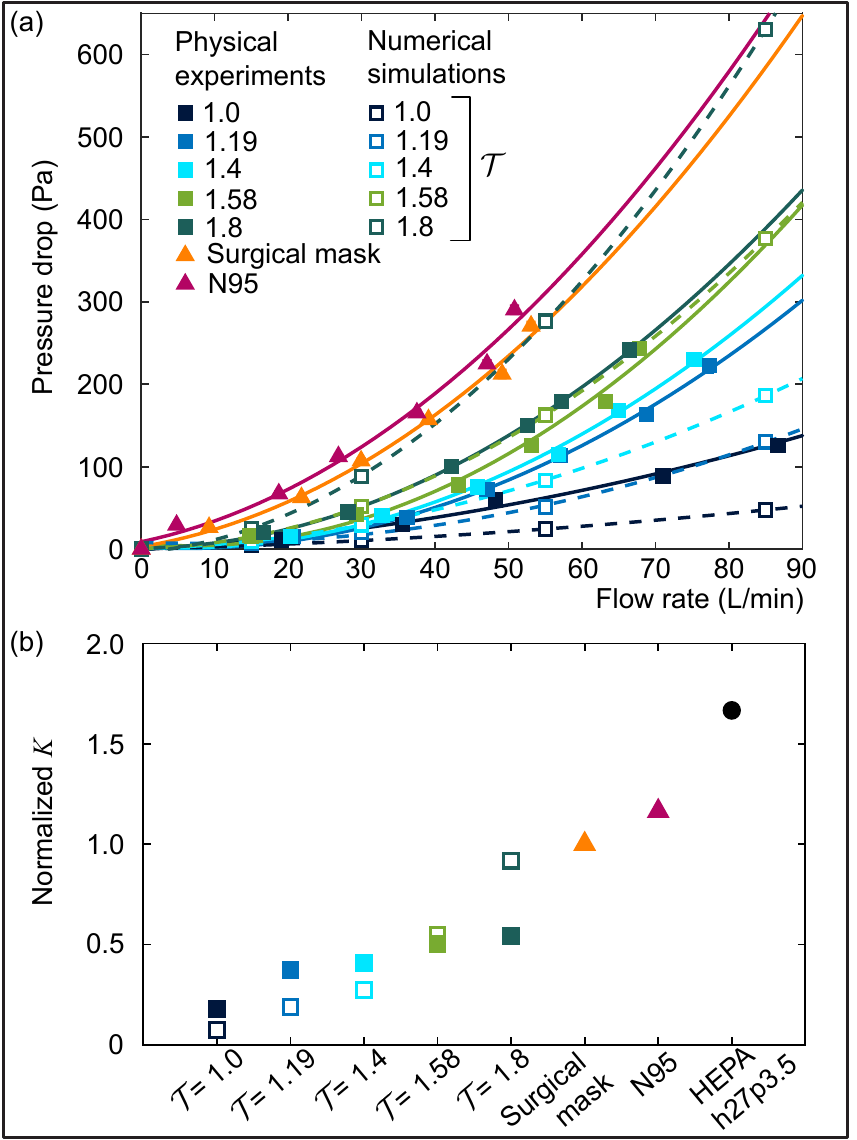}
\caption{(a) Pressure drop across filters with different tortuosity values, ${\cal T}$. The corresponding measurements for the surgical mask, N95 and HEPA filters are included for reference \cite{del2002air}. (b) Pressure loss coefficient $K$, normalized with respect to that of the surgical mask. }
\label{Fig_Pressuretest}
\end{figure}

\begin{figure*}[t]
\centering
\includegraphics[width=1\textwidth]{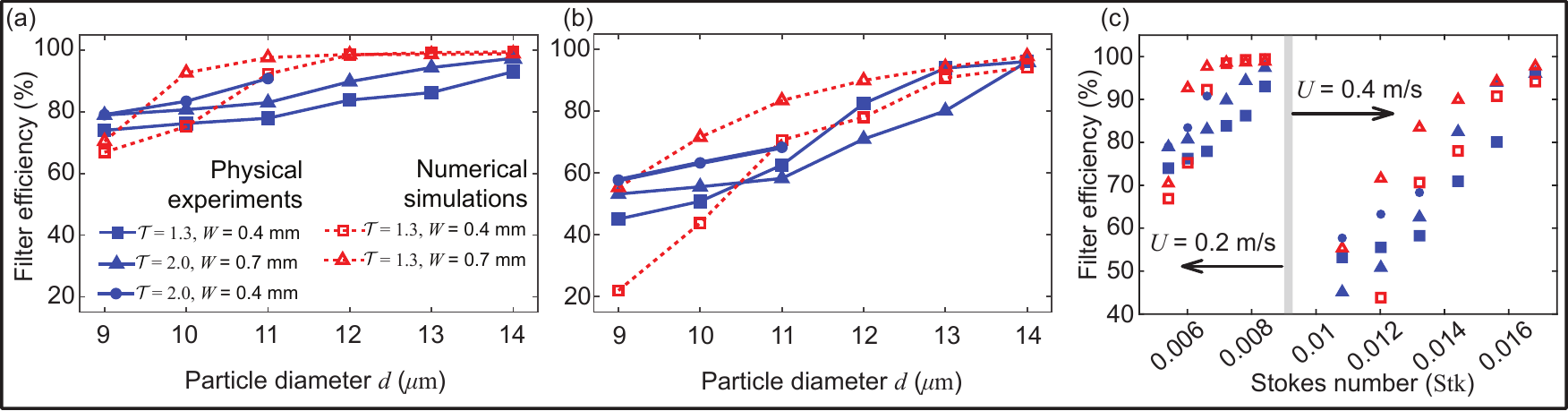} 
\caption{Particle filtering for a mean velocity of (a) 0.2 m/s and (b) 0.4 m/s; experiments include double cartridges. (c) Deposition efficiency versus Stokes number.}
\label{efficiency}
\end{figure*}

\subsection*{Animal nasal cavity analysis} 
We first characterized the nasal cavity structure of 9 distinct mammal species (see Methods \& Materials). Fig. \ref{Fig_animal}a,b illustrates cross-sectional views of nasal cavities of selected adult mammals. These images are anatomical or CT-scanned photos in a plane perpendicular to a line connecting the nostril and the nasopharynx, i.e., coronal view. The air pathway has a complicated labyrinth-like structure; the inhaled and exhaled air takes tortuous paths. The actual air pathway follows a complex 3D trajectory in the connected ways of the coronal and sagittal planes induced by the changes in direction and local gap thickness. This contributes to most olfactory airflow departing from the main respiratory airflow path in the sagittal plane to the dorsal side in the coronal plane \cite{craven2010fluid}. We particularly explored  the geometric properties in the coronal plane to understand how odorant particles can be captured by the olfactory mucosa on the nasal wall.

Basic assessment is done by quantifying and comparing features of nasal cavity geometries, including the gap thickness, $W$, the radius of curvature, $\kappa^{-1}$, and the tortuosity, ${\cal T}$, which are illustrated in Fig. \ref{Fig_animal}c,d. Here, the tortuosity is defined as the ratio of the arc length of a  branch, $L_\mathrm{arc}$, to the shortest Euclidean distance between two points, $L$. We skeletonized the cross-sectional images and defined a branch as connected lines between neighboring pixels to determine ${\cal T}$; see \blue{\textit{SI Appendix}, section A} for details. 

In general, the nasal cavities of large animals, except for humans, have large characteristic length scales, namely, the radius of curvature and gap thickness, which follow an allometric relation $\kappa^{-1} \approx 1.14 \, W^{0.98}$ or $\kappa W^{0.98}\approx 1.14$. However, the nasal structure of humans does not adhere to this trend; it has a comparatively much larger radius of curvature to its gap thickness. The distinct geometric features shared by the animals and not by humans may contribute mechanically to their superior sense of smell. The nasal cavity of pigs shows that the air passage swirls in a spiral fashion (Fig. \ref{Fig_animal}a), which produces the highest tortuosity among the other animals. Comparatively small animals such as marsupials and rodents, Potoroo with longer curvy branches, exhibit the second-highest tortuosity. Other animals also have similar complex nasal structures, but shorter and more straight branches resulting in comparatively low tortuosity. The air is locally connected through various routes within nasal networks, resulting in a broad range of tortuosity measured between 1.2 and 3.3 with a significant deviation. The method of measuring the tortuosity in this study is limited because it does not reflect the complex route of airflow. Instead, by measuring the tortuosity of the branches, the pathway of odorant particles can be inferred. 
Consequently, the tortuosity on the coronal plane approximately captures the complexity of the structure of the nasal bones of animals and provides a guide for the design of synthetic filters.

\subsection*{Basic filter geometry} 
Taking into account the value of the basic parameters obtained in the animal noses, we designed test filters consisting of arrays of conduits containing five levels of tortuosity, ${\cal T}$, following the allometric relationship shown in Fig. \ref{Fig_animal}c. Fig. \ref{tau} illustrates a straight (${\cal T}=$ 1.0) and tortuous (${\cal T}=$ 1.4, 1.8 and 2) channels with square cross-sections. The conduits are tortuous with sharp 90$^\circ$ turns and side walls of $W=1$ or $W=0.4$ mm. The range of tortuosity set, ${\cal T}\in [1,\ 2.0]$, covers most animal's nasal characteristics except for pigs.

\subsection*{{Filter pressure-drop}}
The tortuous filters are first tested for the pressure drop, $\Delta P$, across 60 mm circular cross-section cartridges of 10 mm thickness containing 312 conduits arranged regularly within a 50 mm diameter region. Note that the geometry of the conduits allows for compact packing that minimizes material requirement and pressure drop (see an example in Fig. \ref{tau}c). However, cartridges with an optimized layout are not needed here; a single conduit arrangement in an aligned pattern is better than typical filters such as N95, surgical mask, and HEPA.  As expected, the configurations with a higher level of tortuosity produce higher pressure drop for a given flow rate due to the local loss induced by the channel bends and longer pathway; see Fig.~\ref{Fig_Pressuretest}a. Surgical mask, N95, and HEPA filters are used as reference cases to compare the relative pressure of the proposed filters. 

The bulk pressure loss coefficient, $K$, of the various filters can be characterized as $\Delta P/\rho g = K U^2/2g$. Here, $\rho$ is the air density, $g$ is the gravitational constant, and $U$ is the averaged incoming airspeed. The pressure drop and characteristic coefficient $K$ for each filter are shown in Fig.~\ref{Fig_Pressuretest}b.  The new filters have a substantially lower pressure coefficient than the N95, surgical mask, and HEPA filters. In particular, the highest pressure coefficient of the filter cartridge  is still less than half of those of the surgical mask; $K({\cal T} =1.8) \approx 0.5 K_\mathrm{surgical}$. It indicates the direct benefit of these filters when used as respirators, and the reduced likelihood of air leakage from the boundaries of commercial masks. Complementary numerical simulations of filter cartridges performed for four flow rates using averaged Navier Stokes equations are shown with open symbols in Fig. \ref{Fig_Pressuretest}b.

\subsection*{{Particle capture performance}} 
The particle capture efficiency of the filters as a function of flow velocity and tortuosity is illustrated in Fig. \ref{efficiency}. The higher filtering efficiency is achieved by both decreasing freestream velocity and increasing filter tortuosity. A comparatively greater increase in efficiency for smaller particles (i.e., diameter $d \leq 12~ \mu$m) is achieved with reduced inflow velocity;  increasing tortuosity from 1.3 to 2, produces a small increment in efficiency for all particle sizes as shown in Fig.~\ref{efficiency}a. Fig. \ref{efficiency}c shows the filtering efficiency versus the particle Stokes number, $\mathrm{Stk}$. {Here, $\mathrm{Stk} = d \rho_{p} U /(18 \mu)$ is the Stokes number and $\rho_{p}$ is the density of particle and $\mu$ is the air dynamic viscosity.} Two critical $\mathrm{Stk}$ values are observed for inflow velocities $U = 0.2$ and 0.4 m/s, where the typical positive correlation between deposition efficiency and Stokes number \citep{verjus2016critical,nicolaoustokes,rader2008transport} is followed for different particle sizes; however, this does not hold for the change in inflow velocity. This can be explained by the lower tangential rebound angle for higher inflow velocity caused by a much lower coefficient of restitution \citep{fan1995analysis}, demonstrating that both rebound angle after a collision and $\mathrm{Stk}$ effects are crucial during the particle trapping process within tortuous passages.



The particle capturing simulations in single tortuous pathways of tortuosity 1.3 and 2 agree with the trends from the experimental results. The higher velocity of airflow shows lower capturing efficiency at smaller diameters, whereas above 13 $\mu$m particles, the efficiency is close to 100\% for both tortuosities. Tortuosity increase (turns in the pathway geometry) showed enhanced capturing efficiency.

\vspace{2mm}

\section*{Discussion} 
Underlying bioinspiration of our filter design comprises adopting the well-evolved morphological nasal structures seen in high-olfactory animals, that assist in efficient capture of fine particles embedded in inhaled air onto the olfactory epithelium while maintaining a low pressure drop, thereby allowing easy breathing. The newly-engineered filter system aims to achieve similar goals; high particle capturing efficiency at relatively low-pressure drop. The CT-scanned images of 9 mammal nasal structures showed very tortuous intranasal air paths, quantified by the tortuosity parameter $\cal T$ that varied between 1.2 and 3.3, and the radius of curvature  was linearly proportional to the gap thickness. Based on these basic morphological relationships in animals, we designed filters with simplified geometries and manufactured them using home-use and professional 3D printers. Laboratory tests showed that the pressure drop across the filter increases with tortuosity owing to the additional curves traversed by the incoming air; however, this pressure gradient requirement was in fact almost two times smaller than that of commercial filters (surgical, N95, and HEPA), despite the arrangements of the tortuous conduits being not efficiently packed. The filtration capability can be adjusted easily by changing the tortuosity levels and adding filter layers, if needed. Pressure drop can also be reduced with efficient conduit packing.

Our novel filter design can be useful for many industrial and biomedical applications in which high collection efficiency and low-pressure drop are crucial. For example, many have  experienced a shortage of personal protective devices and medical ventilators during the peak of the pandemic in 2020. At a small scale, our 3D-printed model can replace other mask filters to reduce the pressure drop while supporting high particle capturing rates. It should be also be noted that this bioinspired filter concept is also significantly scalable for various applications that may demand industry-level air filtering.


\subsection*{Animal nasal analysis using CT scan images}
CT scan images for the pig were taken using Toshiba Aquilion 16-slice CT-Scanner, which provides 0.5 mm slice resolution. 
The Dasyurus hallucatus (Northern Quoll),  Caluromys philander (Bare-tailed woolly opossum),  Petauroides volans (greater glider), and Potorous tridactylus (Long-nosed potoroo)'s CT scan images were acquired by Dr.~Macrini \cite{macrini2012comparative}. For the rabbit, the dorsal view of histological section was acquired from Casteleyn et al.~\cite{casteleyn2010nalt}. The high-resolution magnetic resonance image of the dog was obtained from Craven et al.~\cite{craven2009development}. Guinea pig's CT scan images were obtained from Ms.~Jeri Rodgers \cite{rodgers2012cavia}.
The nasal cavity portion from the coronal view of nasal images obtained from MATLAB impage processing toolbox was processed as a black and white binary image to measure the gap thickness, curvature, and tortuosity using ImageJ software. The measurement method of each characteristic is described in the \textcolor{blue}{\textit{SI Appendix}, section A}.

\subsection*{Design and 3D-printing of mask filters}
We utilized key common features of the nasal structure of mammalian turbinates. This structure compartmentalizes the incoming airflow into several tortuous channels in a distinct way that maximizes particle capture. Following a similar relation between the tortuosity and channel width, we design 3D mask structures. The designed filters (${\cal T} \in [1, 1.8]$) were sliced through Cura software (Ultimaker Ltd). The print settings for slicing are set as 0.2 mm for the layer height, 0.4 mm for the line width, 100 \% for the infill percentage, and 0.8 mm for the wall thickness. The filters were 3D printed by using Ultimaker s5 (Ultimaker Ltd). For the highest tortuous filter ${\cal T} = 2.0$, Shapeways corp. used the ProJet® MJP 3500 (3D System Ltd) printer with the VisiJet\textregistered  M3 Crystal material.


\subsection*{Experimental set-up for pressure drop measurements}

A piston was used to control the air flow through the filter, while measuring the pressure drop across the filter. The piston setting consisted of three parts. A 2-in pipe size of the clear rigid pipe (49035K48, McMaster-Carr Supply Co.) was used for the inlet part. A 2-pipe female socket connector (9161K46, McMaster-Carr Supply Co.) is used for the middle part where the filter cartridge is located. The last outlet part is composed with a 2-in pipe size of clear rigid pipe (49035K48, McMaster-Carr Supply Co.) and piston. To construct the piston, a 3-in pipe size of the clear rigid pipe (49035K49, McMaster-Carr Supply Co.) was used for the inner piston, and a 4-in pipe size of clear rigid pipe (49035K51, McMaster-Carr Supply Co.) was used for the outer surrounding inner piston. In addition, a straight reducer (4880K018, McMaster-Carr Supply Co.) was used to connect 2-in pipe size and 4-in pipe size of clear rigid pipe. 

Small holes upwind and downwind the filter cartridge were used to measure pressure drop. A  90$^{\circ}$  anemometer was connected to the pressure sensor and Data Acquisition Card (DAQ Card, DAQ Systems by NI™). The analog input voltage data was measured through the DAQ Card was converted to pressure through a calibration curve following standard procedure (see \textcolor{blue}{\textit{SI Appendix}, section B}). Also, the speed of airflow versus piston speed was measured by using laser sheets, fog particles, and a high-speed camera. The fog particle was injected into the left side of the filter, then it was inhaled to the right side by the piston moving. High-speed videos were used for velocimetry using PIVlab software as a complement.

\subsection*{Experimental set-up for particle capturing efficiency}

The particle capturing efficiency of the designed filters was experimentally measured in a 500 mm long, 40 mm wide, and 40 mm high wind tunnel located in the Renewable Energy and Turbulent Environment Laboratory at the University of Illinois. A Sunon Fans 12V DC brushless fan is installed upwind of the inlet followed by a contraction section with an area ratio of 25:1. A flow straightener was installed near fans and another was installed at the beginning of the channel to ensure flow uniformity. The fan generates volumetric air flows up to 24 cubic feet per minute (CFM). A single ADG-SK508 compression type nozzle is used to generate particles  of diameters ranging from 8 $\mu$m to 15 $\mu$m, which approximately mimics the particle condition of human breathing \citep{bake2019exhaled}. Aqueous chlorophyll solution of about $2.4 \%$ by volume was used as the fluid feeding into the nebulizer to increase particle light reflectivity to a 532 $\mu$m wavelength laser. This  allowed for particle sizing and distribution.

Two field of views (FOVs) of interest are selected with their center located 40 mm and 80 mm upwind and downwind the filter. The incoming and filtered particle size and shape distributions were acquired across a streamwise plane using a low-speed, high-resolution planar PIV system from TSI. Two cameras both equipped with 25 mm, F 2.8 LOWA ultra macro-lens with 5X magnification, were used to interrogate two 3.87 mm $\times$ 2.90 mm field of views (FOVs). The FOVs were located at the center of the wind tunnel in both spanwise and wall normal direction. The FOVs were illuminated from a 250 mJ/pulse Quantel double-pulsed laser. Five sets of 100 image pairs were collected at an acquisition frequency of 2.4 Hz, using a pair of 8MP (3320 $\times$ 1560 pixels) CCD cameras with 16 bit frame-straddle.

\subsection*{Particle Size Analysis}
The iterative Laplacian of Gaussian (LoG) filtering is used for particle sizing. Generalized LoG filter, as introduced in Kong et al. \citep{kong2013generalized}, was used to detect the blobs in the images. The two-dimensional LoG kernel is generated by applying the Laplacian operator $\nabla^{2}$ in the Gaussian function $G(x, y ; \sigma)$, resulting the following kernel: 
\begin{equation}
\nabla^{2} \mathrm{G}(x, y; \sigma)=\frac{x^{2}+y^{2}-2 \sigma^{2}}{\pi \sigma^{4}} \exp \left(-\frac{x^{2}+y^{2}}{2 \sigma^{2}}\right).
\label{LoG}
\end{equation}
The LoG kernel is then convoluted with the blob-like structures to return a fitting score. This process is performed in an iterative manner with increasing $\sigma$ LoG kernels, where the best fit scale of $\sigma_{0}$ is determined when the fitting score converges to a local maximum, the radius of particle is then obtained as $r = \sqrt{2}\sigma_{0}$ \citep{wang2020automated}. This algorithm is applied to each blob-like structure in the image with an adaptive threshold to filter out the unfocused outliers allowing accurate particle size distribution from the PIV images.

\subsection*{Computational Model} 

The \textit{in silico} geometrical setup to track the pressure drop across a filter consists of short upwind and downwind conduits with the circular filter (extracted from the filter stereolithography files) being placed in between. To explore the transport mechanism in more detail, we also reconstructed a single air passage pathway ($\approx$300 of which placed in a circular grid assembly would comprise one complete filter prototype) for each of the tortuous designs and simulated the inhaled airflow and particle transport therein.


\subsection*{Airflow Simulation} 
The numerical simulations in the meshed filter domains considered four airflow rates, viz.~15, 30, 55, and 85 L/min. The low flow rate, even in tortuous pathways, is dominated by viscous-laminar steady-state flow physics \cite{basu2020scirep,inthavong2019cb,zhang2019it,basu2018ijnmbe,farzal2019ifar,kimbell2019lsm}; the higher flow rates trigger separation from the tortuous walls, resulting in flow fluctuations. We have replicated such regimes through Large Eddy Simulation (LES), accounting for the small fluctuations with a sub-grid scale Kinetic Energy Transport Model \cite{baghernezhad2010jt}. The computational scheme employed a segregated solver, with pressure-velocity coupling and second-order upwind spatial discretization. We monitored solution convergence by minimizing the mass continuity and velocity component residuals, and through stabilizing the mass flow rate and static pressure at the airflow outlets. 
The density and dynamic viscosity of inhaled air were set to $1.204$ kg/m$^3$ and $1.825\times10^{-5}$ kg/m$\,$s.

\subsection*{Particle capturing efficiency}
Particle dynamics against the ambient airflow passing through the filter were tracked by Lagrangian-based discrete phase inert transport simulations with the localized deposition and clustering along the filter walls obtained through numerically integrating the transport equation:
\begin{equation}
  \frac{d\mathbf{u}_p}{dt} = \frac{18\mu}{d^2\rho_p} \frac{C_D Re}{24} (\mathbf{u}-\mathbf{u}_p) + \textbf{g}\left(1- \frac{\rho}{\rho_p}\right) + \textbf{F}_\textbf{B}.
\end{equation}
Here $\mathbf{u}_p$ represents the particle velocity, $\mathbf{u}$ is the airflow field velocity, $\rho$ and $\rho_p$ respectively are the density of inhaled air and the particle material density, $\mathbf{g}$ is the gravitational acceleration, and $\mathbf{F_B}$ accounts for any other additional body forces per unit particle mass (e.g., Saffman lift force which is exerted by a typical flow-shear field on small particulates transverse to the airflow direction). The term $18\mu\, C_D\, Re\,(\mathbf{u}-\mathbf{u}_p)/24(d^2 \rho_p)$ quantifies the drag force contribution per unit particle mass, with $C_D$ representing the drag coefficient, $d$ representing the particle diameter, and $Re$ is the relative Reynolds number.
Also, the particle size range is considered large enough to neglect Brownian motion effects on their dynamics.

To replicate the physical conditions of dehydrated airborne particles in the numerical simulations, the particle material density was kept at 1100 kg/m{$^3$}. Trap boundary condition, whereby a particle motion would cease when it reaches the cells adjacent to the geometry surface, is not applied on all inner walls, but on selected walls with comparatively low wall shear stress. Particles are assumed to shear away and separate from the walls with higher wall shear values. Details of the numeric protocol (based off published findings, e.g.~\cite{haron2012coefficient,fan1995analysis}) can be found in \textcolor{blue}{\textit{SI Appendix}, section D}.

\section{Acknowledgement}
{J.Y. and S.J. acknowledge funding support from the National Science Foundation (NSF) grant no.~CBET-2028075. L.P.C. acknowledges funding support from the NSF grant no.~CBET-2028090. S.B. acknowledges funding support from the NSF grant no.~CBET-2028069. }

\bibliography{refv2}

\end{document}